\documentclass[12pt]{article}

\usepackage{epsfig,latexsym}
\usepackage{amsmath}

\newcommand{\join}{\text{\textcircled{{\footnotesize 1}}}}
\newcommand{\cojoin}{\text{\textcircled{{\footnotesize 0}}}}

\newcommand{\qed}{\hfill $\Box$}

\newtheorem{fact}{Fact}
\newtheorem{theo}{Theorem}
\newtheorem{lemm}{Lemma}

\newtheorem{obse}{Observation}

\def\inst#1{$^{#1}$}

\title{Maximum Weight Independent Sets for ($P_7$,Triangle)-Free Graphs in Polynomial Time}

\author{Andreas Brandst\"adt\inst{1}
\and
Raffaele Mosca\inst{2}
}

\begin{document}



\maketitle

\begin{center}
{\footnotesize
 \inst{1} Institut f\"ur Informatik, Universit\"at Rostock, D-18051 Rostock, Germany.\\
\texttt{ab@informatik.uni-rostock.de}

\inst{2} Dipartimento di Economia, Universit\'a degli Studi "G. D'Annunzio", Pescara 65121, Italy.\\
\texttt{r.mosca@unich.it}
}
\end{center}

\begin{abstract}
The Maximum Weight Independent Set (MWIS) problem on finite undirected graphs with vertex weights asks for a set of pairwise nonadjacent vertices of maximum weight sum. MWIS is one of the most investigated and most important algorithmic graph problems; it is well known to be NP-complete, and it remains NP-complete even under various strong restrictions such as for triangle-free graphs. Its complexity was an open problem for $P_k$-free graphs, $k \ge 5$. Recently, Lokshtanov et al. \cite{LokVatVil2014} proved that MWIS can be solved in polynomial time for $P_5$-free graphs, and Lokshtanov et al. \cite{LokPilvan2015} proved that MWIS can be solved in quasi-polynomial time for $P_6$-free graphs. It still remains an open problem whether MWIS can be solved in polynomial time for $P_k$-free graphs, $k \geq 6$ or in quasi-polynomial time for $P_k$-free graphs, $k \geq 7$. Some characterizations of $P_k$-free graphs and some progress are known in the literature but so far did not solve the problem. In this paper, we show that MWIS can be solved in polynomial time for ($P_7$,triangle)-free graphs. This extends the corresponding result for 
($P_6$,triangle)-free graphs and may provide some progress in the study of MWIS for $P_7$-free graphs. 
\end{abstract}

Keywords: Graph algorithms; Maximum Weight Independent Set problem; $P_7$-free graphs; triangle-free graphs; polynomial time algorithm; anti-neighborhood approach.

\section{Introduction}

An {\em independent set} (or {\em stable set}) in a graph $G$ is a subset of pairwise nonadjacent vertices of $G$. An independent set in a graph $G$ is {\em maximal} if it is not properly contained in any other independent set of $G$.

Given a graph $G$ and a weight function $w$ on $V(G)$, the Maximum Weight Independent Set (MWIS) problem asks for an independent set of $G$ with maximum weight. Let $\alpha_w(G)$ denote the maximum weight of an independent set of $G$. The MWIS problem is called {\em MIS problem} if all vertices $v$ have the same weight $w(v) = 1$.

The MIS problem ([GT20] in \cite{GareyJohn1979}) is well known to be NP-complete \cite{Karp1972}. While it is solvable in polynomial time for bipartite graphs (see e.g. \cite{AhuMagOrl1993,DesHak1970,GroLovSch1988}), it remains NP-hard even under various strong restrictions, such as for triangle-free graphs \cite{Polja1974}. 

The following specific graphs are subsequently used. A $P_k$ has vertices $v_1,v_2,\ldots,v_k$ and edges $v_jv_{j+1}$ for $1 \le j < k$. A $C_k$ has vertices $v_1,v_2,\ldots,v_k$ and edges $v_jv_{j+1}$ for $1 \le j \le k$ (index arithmetic modulo $k$). A $K_{\ell}$ has $\ell$ vertices which are pairwise adjacent. A $K_3$ is also called {\em triangle}.
A {\em claw} has vertices $a,b,c,d$ and edges $ab,ac,ad$. A $S_{i,j,k}$ is the graph obtained from a claw by subdividing respectively its edges into $i$, $j$, $k$ edges (e.g., $S_{0,1,2}$ is a $P_4$, $S_{1,1,1}$ is a claw).

For a given graph $F$, a graph $G$ is {\em $F$-free} if no induced subgraph of $G$ is isomorphic to $F$. If for given graphs $F_1,\ldots,F_k$, $G$ is $F_i$-free for all $1 \le i \le k$ then we say that $G$ is {\em $(F_1,\ldots,F_k)$-free}.

Alekseev \cite{Aleks1983,Aleks2004/2} proved that, given a graph class ${\cal{X}}$ defined by forbidding a finite family $\cal{F}$ of induced graphs, the MIS problem remains NP-hard for the graph class ${\cal{X}}$  if each graph in $\cal{F}$ is not a $S_{i,j,k}$ for some index $i,j,k$. Various authors \cite{FaeOriSta2011,Minty1980,NakTam2001,NobSas2015,Sbihi1980} proved that MWIS can be solved for claw-free (i.e., $S_{1,1,1}$-free) graphs in polynomial time even by fast solution methods. Lozin and Milani\v c \cite{LozMil2008} proved that MWIS can be solved for fork-free graphs (i.e., $S_{1,1,2}$-free graphs) in polynomial time $-$ Alekseev \cite{Aleks1999,Aleks2004/1} previously proved a corresponding result for the unweighted case. Lokshtanov et al. \cite{LokVatVil2014} recently proved that MWIS can be solved for $P_5$-free graphs (i.e., $S_{0,2,2}$-free graphs) in polynomial time.

Then by the above, the class of $P_6$-free graphs is one of three minimal graph classes, defined by forbidding a single connected subgraph, for which the computational complexity of MIS is an open question. However Lokshtanov et al. \cite{LokPilvan2015} recently proved that MWIS can be solved for $P_6$-free graphs in quasi-polynomial time $n^{{\cal O}(\log^2n)}$. Then $-$ also referring to the conclusions of \cite{LokVatVil2014} $-$ a natural open problem is to establish if MWIS can be solved for $P_k$-free graphs in polynomial time for $k \geq 6$ or in quasi-polynomial time for $k \geq 7$, even though some characterizations of $P_k$-free graphs and some progress are known in the literature; see e.g. respectively \cite{Bacso2009,BacTuz1990/1,BacTuz1990/2,CamSch2015,Dong1999,HofPau2010,Tuza1980} and \cite{BraHoa2005,GerHerLoz2003,Karth2015,LozRau2003,Mosca2008,Mosca2009,Mosca2010,MoscaDMTCS,MoscaIPL}.

In this manuscript, we show that MWIS can be solved for ($P_7$,triangle)-free graphs in polynomial time. This extends the corresponding result for ($P_6$,triangle)-free graphs and may provide some progress in the study of MWIS for $P_7$-free graphs. While MWIS remains NP-hard for triangle-free graphs, it is an open problem whether MWIS remains NP-hard on ($P_k$,triangle)-free graphs for some $k > 7$.

Let us recall that the class of ($P_6$,triangle)-free graphs has been studied in various papers, see e.g. \cite{BraKleMah2004,Brand2002,LiuZho1994,Mosca1999,RanSchTew2002}, where various structure properties have been introduced and often applied to solve MWIS for such graphs.
In particular, Brandst\"adt et al. \cite{BraKleMah2004} showed that ($P_6$,triangle)-free graphs have bounded clique-width, which implies that a large class of NP-hard problems (including MWIS) can be very efficiently solved for such graphs.
Let us mention that on the other hand, $P_7$-free bipartite graphs $-$ and thus ($P_7$,triangle)-free graphs $-$ have unbounded clique-width \cite{LozVol2008}.

Our approach is based on a repeated application of the {\em anti-neighborhood approach} which is based in detail on the easy observation that for any vertex $v$ of a graph $G=(V,E)$ and its neighborhood $N(v)$ in $G$, one has 
$$\alpha_w(G) = \max \{\alpha_w(G[V \setminus \{v\}]),\alpha_w(G[V \setminus N(v)])\}.$$ 
That allows, by detecting an opportune sequence of vertices, to split and to finally reduce the problem to certain instances of bipartite subgraphs, for which the problem can be solved in polynomial time \cite{AhuMagOrl1993,DesHak1970,GroLovSch1988}. In particular as a corollary one obtains: For every ($P_7$,triangle)-free graph $G$ there is a family ${\cal S}$ of subsets of $V(G)$ inducing bipartite subgraphs of $G$, with ${\cal S}$ detectable in polynomial time and containing polynomially many members, such that every maximal independent set of $G$ is contained in some member of ${\cal S}$. That seems to be harmonic to the result of Pr\"omel et al. \cite{ProSchSte2002} showing that with ``high probability'', removing a single vertex in a triangle-free graph leads to a bipartite graph.

\section{Notations and preliminary results}

For any missing notation or reference let us refer to \cite{BraLeSpi1999}.

Let $G$ be a graph and let $V(G)$ (respectively, $E(G)$) denote the vertex set (respectively, the edge set) of $G$. 
For $U \subseteq V(G)$, let $G[U]$ denote the subgraph of $G$ induced by $U$. Throughout this paper, all subgraphs are understood as induced subgraphs. 

For $v \in V(G)$, let $N(v) = \{u \in V(G) \setminus \{v\} : uv \in E(G)\}$ be the {\em open neighborhood} of $v$ in $G$, let $N[v]=N(v) \cup \{v\}$ be the {\em closed neighborhood} of $v$ in $G$, and let $A(v)=V(G) \setminus N[v]$ be the {\em anti-neighborhood} of $v$ in $G$. If $u \in N(v)$ ($u \notin N(v)$, respectively) we say that {\em $u$ sees $v$} ({\em $u$ misses $v$}, respectively).  

For $v \in V(G)$ and $U \subseteq V(G)$, with $v \not \in U$, let $N_U(v) = N(v) \cap U$. 

For $U,W \subseteq V(G)$, with $U \cap W = \emptyset$, $U$ has a {\em join} (a {\em co-join}, respectively) to $W$, denoted by $U \join W$ ($U \cojoin W$, respectively), if each vertex in $W$ is adjacent (is nonadjacent, respectively) to each vertex in $U$.

For $v \in V(G)$ and $U \subseteq V(G)$, with $v \not \in U$, $v$ {\em contacts} $U$ if $v$ is adjacent to some vertex of $U$; 
$v$ {\em dominates} $U$ if $v$ is adjacent to all vertices of $U$, that is, $\{v\} \join U$ ($v \join U$ for short); 
$v$ {\em misses} $U$ if $v$ is non-adjacent to all vertices of $U$, that is, $\{v\} \cojoin U$ ($v \cojoin U$ for short).

A {\em component of $G$} is a maximal connected subgraph of $G$. 
The {\em distance} $d_G(u,v)$ of two vertices $u,v$ in $G$ is the number of edges of $G$ in a shortest path between $u$ and $v$ in $G$.

For a subgraph $H$ of $G$ and $k \ge 0$, a vertex $v \notin V(H)$ is a {\em $k$-vertex for $H$} (or {\em of $H$}) if it has exactly
$k$ neighbors in $H$. {\em $H$ has no $k$-vertex} if there is no $k$-vertex for $H$.

The following result concerning M(W)IS for bipartite graphs is well known:

\begin{theo}[\cite{AhuMagOrl1993,DesHak1970,GroLovSch1988}]\label{MWSbipgrpoltime}
Let $B$ be a bipartite graph with $n$ vertices.
\begin{enumerate}
\item[$(i)$]  MWIS $($with rational weights$)$ is solvable for $B$ in time ${\cal O}(n^4)$ via linear programming
              or network flow.
\item[$(ii)$] MIS is solvable for $B$ in time ${\cal O}(n^{2.5})$.   
\end{enumerate}
\end{theo}

A graph $G$ is {\em nearly bipartite} if, for each $v \in V(G)$, the subgraph $G[A(v)]$ induced by its anti-neighborhood is bipartite. Obviously one has:
$$\alpha_w(G) = \max_{v \in V(G)} \{w(v)+\alpha_w(G[A(v)])\}$$

Thus, by Theorem \ref{MWSbipgrpoltime}, the MWIS problem (with rational weights) can be solved in time ${\cal O}(n^5)$ for nearly bipartite graphs.

Subsequently, the following lemma will be useful:

\begin{lemm}\label{P7K3C5frnearlybip}
Connected $(P_7$,triangle,$C_5)$-free graphs are nearly bipartite.
\end{lemm}

{\bf Proof.} Let $G=(V,E)$ be a connected $(P_7$,triangle,$C_5)$-free graph. Assume to the contrary that $G[A(v)]$ is not bipartite for some vertex $v$, i.e., $G[A(v)]$ contains an odd cycle. Then, since $G$ is ($P_7$,triangle,$C_5$)-free, $G[A(v)]$ contains a $C_7$, say $C$. Since $G$ is $P_7$-free, $C$ has no 1-vertex; since $G$ is (triangle,$C_5$)-free, $C$ has no 3-vertex; since $G$ is triangle-free, $C$ has no $k$-vertex for $4 \le k \le 7$. Since $G$ is (triangle,$C_5$)-free, the neighbors of 2-vertices have distance 2 in $C$. Since $G$ is $P_7$-free, 0-vertices are nonadjacent to 2-vertices, a contradiction since $G$ is connected and $v$ is a 0-vertex of $C$. Thus, for every $v \in V$, $G[A(v)]$ is bipartite. 
\qed

Moreover, since $G$ is ($P_7$,triangle)-free, we obviously have:
\begin{obse}\label{disjointC5s}
There are at least two edges between any two vertex-disjoint $C_5$'s in~$G$.  
\end{obse}

\section{MWIS for ($P_7$,triangle)-free graphs}

In this section we are going to show that MWIS can be solved in polynomial time for ($P_7$,triangle)-free graphs. Since by Lemma \ref{P7K3C5frnearlybip}, we are done with $(P_7$,triangle,$C_5)$-free graphs, from now on let $G$ be a connected ($P_7$,triangle)-free graph containing a $C_5$. 

Using the anti-neighborhood approach, it is sufficient to show that for every vertex, say $c \in V(G)$, and for every component of its anti-neighborhood, say $K$ of $G[A(c)]$, MWIS can be solved in polynomial time. Since $G$ is connected, $c$ has a neighbor $d \in N(c)$ contacting $K$. Let 
\begin{enumerate}
\item[ ] $H := V(K) \cap N(d)$ and 
\item[ ] $Z := V(K) \setminus N(d)$. 
\end{enumerate}

Obviously, $\{H,Z\}$ is a partition of $V(K)$. Since $G$ is triangle-free, $H$ is an independent set.
The following observations obviously hold since $G$ is $P_7$-free:

\begin{obse}\label{noHZP3P3}
There is no pair of vertices $h_1,h_2 \in H$ such that $h_1$ and $h_2$ are the endpoints of pairwise disjoint $P_3$'s induced by $\{h_1,z_1,z_2\}$, and $\{h_2,z'_1,z'_2\}$, respectively, such that $z_1,z_2,z'_1,z'_2 \in Z$, $\{z_1,z_2\} \cap \{z'_1,z'_2\} = \emptyset$, with 
$\{h_1,z_1,z_2\} \cojoin \{h_2,z'_1,z'_2\}$. 
\end{obse}

\begin{obse}\label{noHP5Z}
No vertex $h \in H$ is the endpoint of a $P_5$ formed together with four vertices of $Z$.  
\end{obse}

For showing that MWIS can be solved for $K$ in polynomial time, let us consider the following exhaustive cases: 

\subsection{Case 1: $G[Z]$ is bipartite.}

Recall that, if $K$ contains no $C_5$, then, by Lemma \ref{P7K3C5frnearlybip}, MWIS can be solved in polynomial time for $K$.
Thus assume that $K$ contains a $C_5$. Note that, since by assumption, $G[Z]$ is bipartite
and $H$ is an independent set, every $C_5$ $C$ in $K$ is of one of the following two types:

\begin{enumerate}
\item[ ] {\em Type $1$}: $C$ contains one vertex from $H$ and four vertices from $Z$.
\item[ ] {\em Type $2$}: $C$ contains two vertices from $H$ and three vertices from $Z$.
\end{enumerate}

\subsubsection{Case 1.1: $K$ contains no $C_5$ of type 1.}

For any $h \in H$ and for any nontrivial component $T$ of $G[Z]$, with bipartition $T = (U_1, U_2, E')$,
let us say that 
\begin{itemize}
\item[ ] {\em $h$ has a half-join to $T$} if either $N_T(h) = U_1$ or $N_T(h) = U_2$;
\item[ ] {\em $h$ properly one-side contacts $T$} if either $\emptyset \subset N_T(h) \subset U_1$ or
$\emptyset \subset N_T(h) \subset U_2$. 
\item[ ] $T$ is a {\em green component of $G[Z]$} if there is a vertex of $H$ which properly one-side contacts $T$.
\end{itemize}

\begin{fact}\label{fact1}
For every $h \in H$ one has:
\begin{itemize}
\item[$(i)$] if $h$ contacts a nontrivial component of $G[Z]$, say $T$, then either $h$ has a half-join to $T$ or $h$ properly one-side contacts $T$;
\item[$(ii)$] $h$ properly one-side contacts at most one nontrivial component of $G[Z]$.
\end{itemize}
\end{fact}

{\bf Proof.} Fact \ref{fact1} holds since $G$ is ($P_7$,triangle)-free, and since by assumption, $G[Z]$ is bipartite and $K$ contains no $C_5$ of type 1. 
\qed\\

{\bf Case 1.1.1.} $G[Z]$ has no green component.

\begin{lemm}\label{ifnogreencomponent}
If there is no green component of $G[Z]$ then MWIS is solvable in polynomial time for $K$. 
\end{lemm}

{\bf Proof.} 
If $G[Z]$ has no green component then by Fact \ref{fact1}, for each $h \in H$ and for each nontrivial component $T$ of $G[Z]$, either $h$ does not contact $T$ or $h$ has a half-join to $T$. 
Let $H' := \{h \in H: h$ is the endpoint of a $P_3$ formed together with two vertices of $Z\}$.
If $H' = \emptyset$ then $K$ has no $C_5$ of type 2, i.e., by assumption of Case 1, $K$ is $C_5$-free. 
Thus, by Lemma \ref{P7K3C5frnearlybip}, MWIS can be solved in polynomial time for $K$. 

Now assume that $H' \neq \emptyset$. Let '$\ge$' be the following binary relation on $H'$: For any pair $a,b \in H'$, $a \ge b$ if either $a=b$ or $a$ contacts all $P_3$'s formed by $b$ as endpoint and by two vertices from $Z$. Then Observation \ref{noHZP3P3} implies that for any $a,b \in H'$, either $a \ge b$ or $b \ge a$. 

The following argument shows that $\ge$ is also transitive: Assume that $a \ge b$ and $b \ge c$. We claim that this implies $a \ge c$:
Let $c,c',c''$ be a $P_3$ with vertices $c',c'' \in Z$. Then, since $b \ge c$, $bc' \in E$ or $bc'' \in E$. If $bc' \in E$ then,
since $G$ is triangle-free, $bc'c''$ is a $P_3$ contacted by $a$ since $a \ge b$. Thus, $ac' \in E$ or $ac'' \in E$ which
shows that $a$ contacts the $P_3$ $cc'c''$. In the other case, namely when $bc'' \in E$, the argument is similar. This shows that $\ge$ is transitive.

Thus, $\ge$ implies a total order on $H'$, say $H' = \{h_1,\ldots,h_{\ell}\}$, with $h_1 \ge \ldots \ge h_{\ell}$.

Note that, by definition of $h_1$, $G[V(K) \setminus N(h_1)]$ has no $C_5$ of type 2, i.e., $G[V(K) \setminus N(h_1)]$ is $C_5$-free by assumption of Case 1. Then MWIS can be solved for $G[V(K) \setminus N(h_1)]$ in polynomial time by Lemma \ref{P7K3C5frnearlybip}.
Then MWIS can be solved on $K$ by successively solving MWIS in $G[V(K) \setminus N(h_1)]$, in 
$G[(V(K) \setminus \{h_1,\ldots,h_{i-1}\}) \setminus N(h_i)]$ for $i \in \{2,\ldots,\ell\}$, and in $G[V(K) \setminus H']$. Since such graphs are $C_5$-free by construction, as shown above, this can be done in polynomial time by Lemma \ref{P7K3C5frnearlybip}.
This finally shows Lemma \ref{ifnogreencomponent}.
\qed

{\bf Case 1.1.2.} $G[Z]$ has green components.

Let $\{T_1,\ldots,T_k\}$, $k \ge 1$, denote the family of green components of $G[Z]$, and let $H_i:=\{h \in H: h$ properly one-side contacts $T_i\}$ for $i \in \{1,\ldots,k\}$. By Fact \ref{fact1}, we have $H_i \cap H_j = \emptyset$ for $i \neq j$. 

\begin{fact}\label{fact2}
For every $(h_1,\ldots,h_k) \in H_1 \times \ldots \times H_k$, there is an $i \in \{1,\ldots,k\}$ such that for each $j \in \{1,\ldots,k\}$, $j \neq i$,
$h_i$ has a half-join to $T_j$.
\end{fact}

{\bf Proof.} The proof is done by induction on $k$. If $k = 2$, then the assertion follows by Observation \ref{noHZP3P3} and Fact \ref{fact1}.
Then let us assume that the assertion is true for $k-1$ and prove that it is true for $k$. 
Let $(h_1,\ldots,h_k) \in H_1 \times \ldots \times H_k$. By the inductive assumption on $(h_2,\ldots,h_k)$, one can assume without loss of generality that
$h_2$ has a half-join to $T_j$ for every $j > 2$. If $h_2$ has a half-join to $T_1$ then Fact \ref{fact2} is proved. 
Otherwise, by Fact~\ref{fact1}, assume that $h_2 \cojoin T_1$. If for every $j > 1$, $h_1$ has a half-join to $T_j$ then Fact \ref{fact2} is proved. Otherwise, by the inductive assumption on $(h_1,h_3,\ldots,h_k)$, one can assume without loss of generality that $h_3$ has a half-join to $T_1$ and to $T_j$ for every $j > 3$. Note that $h_3$ has a half-join to $T_2$ since otherwise, $h_3$ is the endpoint of a $P_3$ formed together with two vertices of $T_1$, $h_2$ is the endpoint of a $P_3$ formed
together with two vertices of $T_2$, and both $h_2$ and $h_3$ are adjacent to $d$, that is, a $P_7$ arises. Then for each $j \in \{1,\ldots,k\}$, $j \neq 3$, $h_3$ has a half-join to $T_j$. This shows Fact \ref{fact2}.
\qed

For any green component $T$ of $G[Z]$, with $T = (U_1,U_2,E')$, let us define the following notations with respect to $U_1$ (and correspondingly for $U_2$):
\begin{itemize}
\item[ ] $H(U_1) := \{h \in H: \emptyset \subset N_{T}(h) \subset U_1\}$ denotes the set of vertices of $H$ which properly one-side contact $T$ with respect to $U_1$. 
\item[ ] A vertex $h^* \in H(U_1)$ is {\em $U_1$-maximal for $T$} if $N_{U_1}(h^*)$ is inclusion-maximal in $\{N_{U_1}(h): h \in H(U_1)\}$, i.e., 
there is no $h \in H(U_1)$ such that $N_{U_1}(h^*) \subset N_{U_1}(h)$.  
\end{itemize}

\begin{fact}\label{fact3}
Let $T$ be a green component of $G[Z]$ with $T=(U_1,U_2,E')$.
\begin{enumerate}
\item[$(i)$] Let $h_1,h_2 \in H(U_1)$. Then for any $x_1 \in N_{U_1}(h_1) \setminus N_{U_1}(h_2)$ and $x_2 \in N_{U_1}(h_2) \setminus N_{U_1}(h_1)$ we have $N_{U_2}(x_1)= N_{U_2}(x_2)$. 

\item[$(ii)$] If $h^*$ is $U_1$-maximal for $T$, then there exists a subset $Y$ of $U_2$ such that for all $h \in H(U_1) \setminus \{h^*\}$ and for all $x \in N_{U_1}(h) \setminus N_{U_1}(h^*)$ we have $N_{U_2}(x) = Y$. 
\end{enumerate}
\end{fact}

{\bf Proof.}
$(i)$: Suppose to the contrary that there is $y_1 \in U_2$ with $x_1y_1 \in E$ such that $x_2y_1 \notin E$. Then, by the connectedness of $T$, there is $y_2 \in U_2$, $y_2 \neq y_1$, with $x_2y_2 \in E$. Now if $x_1y_2 \notin E$, we get a $P_7$ corresponding to Observation \ref{noHZP3P3}, and if $x_1y_2 \in E$, we get a $P_7$ corresponding to Observation \ref{noHP5Z} which is a contradiction. 

$(ii)$: Let $h$ be any vertex in $H(U_1) \setminus \{h^*\}$,  with $N_{U_1}(h) \setminus N_{U_1}(h^*) \neq \emptyset$. By the $U_1$-maximality of $h^*$, there is a $x^* \in N_{U_1}(h^*) \setminus N_{U_1}(h)$, and thus, by ($i$), $N_{U_2}(x) = N_{U_2}(x^*)$  for all $x \in N_{U_1}(h) \setminus N_{U_1}(h^*)$. Then let us write $N_{U_2}(x ) = Y$ for all $x \in N_{U_1}(h) \setminus N_{U_1}(h^*)$. 

Then let $h'$ be any (possible) vertex in $H(U_1) \setminus \{h^*\}$, with $N_{U_1}(h') \setminus N_{U_1}(h^*) \neq \emptyset$ and with $h' \neq h$, and let us show that for all $x' \in N_{U_1}(h') \setminus N_{U_1}(h^*)$, one has $N_{U_2}(x') = Y$. Let $x'$ be any vertex in $N_{U_1}(h') \setminus N_{U_1}(h^*)$: If $x' \in N_{U_1}(h)$, then the assertion directly follows by the previous paragraph and by definition of $Y$; if $x' \not \in N_{U_1}(h)$, then the assertion follows by ($i$) and by definition of $Y$. This completes the proof of Fact \ref{fact3} $(ii)$.
\qed 

Subsequently we say that a vertex $h \in H$ is {\em maximal for $T_i=(U_{1,i},U_{2,i},E_i)$} if it is $U_{1,i}$-maximal or $U_{2,i}$-maximal as defined above. 

Let us say that a vertex $h \in H$ is a {\em critical vertex of $K$} if 
\begin{itemize}
\item[$(i)$] there is an $i \in \{1,\ldots,k\}$ such that $h$ is maximal for $T_i$, and 
\item[$(ii)$] for each $j \in \{1,\ldots,k\}$, $j \neq i$, $h$ has a half-join to $T_j$. 
\end{itemize}

\begin{fact}\label{fact4}
There is a critical vertex of $K$.
\end{fact}

{\bf Proof.} Let $(h_1^*,\ldots,h_k^*) \in H_1 \times \ldots \times H_k$ such that $h_i^*$ is maximal for $T_i$, for all
$i \in \{1,\ldots,k\}$. Then Fact \ref{fact4} follows by Fact \ref{fact2}. 
\qed  

Finally let us show that in Case 1.1.2, MWIS can be solved in polynomial time for $K$:

\begin{fact}\label{factMWISantineighbcritical}
For any critical vertex, say $h^*$ of $K$, MWIS can be solved in polynomial time for $G[V(K) \setminus N(h^*)]$.
\end{fact}

{\bf Proof.} 
By definition of a critical vertex of $K$, let $T$ be the green component of $G[Z]$, with bipartition $T = (U_1,U_2,E')$, such that $h^*$ is maximal for $T$. 
Let $X = \{h \in H: N_{U_1}(h) = U_1\}$, and let $Y$ be the subset of $U_2$ defined as in Fact \ref{fact3}. Then MWIS can be solved for $G[V(K) \setminus N(h^*)]$ by successively solving MWIS on
\begin{itemize}
\item[$(i)$] $G[V(K) \setminus (N(h^*) \cup N(h))]$ for all $h \in X$; 
\item[$(ii)$] $G[V(K) \setminus (N(h^*) \cup N(y))]$ for all $y \in Y$; 
\item[$(iii)$] $G[(V(K) \setminus (X \cup Y)) \setminus N(h^*)]$.
\end{itemize}

In particular, by Fact \ref{fact3} and since $h^*$ is a critical vertex of $K$, such graphs restricted to their intersection with $Z$ have no green component (as one can easily check). Then steps $(i),(ii)$ and $(iii)$ can be executed in polynomial time by referring to Case 1.1.1. which shows Fact \ref{factMWISantineighbcritical}.
\qed

Using Fact \ref{factMWISantineighbcritical}, MWIS can be solved in polynomial time for $K$ as follows: 

Let us write $H_1 \cup \ldots \cup H_k = \{h_1,\ldots,h_m\}$ (with $k \le m$), such that, according to Fact \ref{fact4}, $h_i$ is a critical vertex of (a connected component of) $G[V(K) \setminus \{h_1,\ldots,h_{i-1}\}]$ for $i \in \{1,\ldots,m\}$.

Then MWIS can be solved for $K$ by successively solving MWIS 
\begin{itemize}
\item[$(i)$] in $G[V(K) \setminus N(h_1)]$ 
\item[$(ii)$] in $G[(V(K) \setminus \{h_1,\ldots,h_{i-1}\}) \setminus N(h_i)]$ for $i \in \{2,\ldots,m\}$,
\item[$(iii)$] in $G[V(K) \setminus (H_1 \cup \ldots \cup H_k)]$. 
\end{itemize}

In particular, steps $(i)-(ii)$ can be executed in polynomial time by the above, and step $(iii)$ can be executed in polynomial time by referring to Case 1.1.1. 

\subsubsection{Case 1.2: $K$ contains a $C_5$ of type 1.}

For any $C_5$ of type 1 in component $K$, say $C$ with vertex set $V(C)=\{v_1, \ldots, v_5\}$ and edges $v_iv_{i+1}$ (index arithmetic modulo 5) such that 
$V(C) \cap H =\{v_5\}$, let us say that $v_5$ is the {\em nail} $h=v_5$ of $C$, and the other vertices of $C$ are the {\em non-nail vertices} of $C$. 
For any such nail $h$, let 
\begin{itemize}
\item[ ] $L(h) := \{z \in Z: z$ belongs to a $C_5$ of type 1 in $K$ with nail $h$, and $zh \notin E\}$. 
\end{itemize}
Note that $v_2,v_3 \in L(h)$.

\begin{fact}\label{fact5}
For any nail $h$, we have:
\begin{itemize}
\item[$(i)$] There is no $C_5$ of type $1$ in $G[V(K) \setminus (N(h) \cup N(v_3))]$, and by symmetry, the same holds for $v_2$ instead of $v_3$. 
\item[$(ii)$] Every $C_5$ $C'$ of type $1$ in $G[V(K) \setminus N(h)]$ contains at least one non-nail vertex $x$ with respect to a $C_5$ of type $1$ with nail $h$ such that $x \in L(h)$.
\end{itemize}

\end{fact}

{\bf Proof.} 
Let $C'$ be a $C_5$ of type 1 in $G[V(K) \setminus N(h)]$, say, with vertex set $V(C')=\{u_1, \ldots, u_5\}$ and edges $u_iu_{i+1}$ (index arithmetic modulo 5), such that $V(C') \cap H =\{u_5\}$. Clearly $h=v_5 \notin V(C')$ and $v_1,v_4 \notin V(C')$ since $V(C') \cap N(h)=\emptyset$. 

We first claim that $\{u_1, \ldots, u_5\} \cap \{v_1, \ldots, v_5\}=\emptyset$: 

If, as in $(i)$, $C'$ is a $C_5$ of type $1$ in $G[V(K) \setminus (N(h) \cup N(v_3))]$ then clearly also $v_2 \notin V(C')$. Moreover, $v_3 \notin V(C')$ since $C'$ contains no neighbor of $v_3$. Thus, in this case, $\{u_1, \ldots, u_5\} \cap \{v_1, \ldots, v_5\}=\emptyset$.

Now assume that $C'$ is a $C_5$ of type $1$ in $G[V(K) \setminus N(h)]$. If $v_2 \in V(C')$ or $v_3 \in V(C')$ then $(ii)$ is fulfilled. Thus also for $(ii)$, we can assume that $v_2,v_3 \notin V(C')$, and the claim is shown. $\diamond$ 

Since by assumption of Case 1, $G[Z]$ is bipartite, we can assume without loss of generality that $v_1,v_3,u_1,u_3$ form an independent set (say, $v_1,v_3,u_1,u_3$ are black), and the same can be assumed for $v_2,v_4,u_2,u_4$ (say, $v_2,v_4,u_2,u_4$ are grey). 

Since $G$ is triangle-free, $u_5$ has at most two neighbors in $C$, and if $|N(u_5) \cap V(C)|=2$ then either $N_C(u_5)=\{v_1,v_4\}$ or $N_C(u_5)=\{v_2,v_4\}$ or $N_C(u_5)=\{v_1,v_3\}$; by symmetry we can assume $N_C(u_5)=\{v_2,v_4\}$. 

If $v_1$ is the only neighbor of $u_5$ in $C$ then $c,d,u_5,v_1,v_2,v_3,v_4$ induce a $P_7$ in $G$, which is a contradiction, and an analogous argument holds if $v_4$ is the only neighbor of $u_5$ in $C$.
Thus, for $|N(u_5) \cap V(C)|=1$, by symmetry, only $N_C(u_5)=\{v_2\}$ has to be considered.  

We show $(i)$ and $(ii)$ by analyzing the following cases:  

{\bf Case A}: $N_C(u_5)=\{v_1,v_4\}$.

Since $G$ is triangle-free, $\{v_1,v_4,u_1,u_4\}$ is an independent set.

Since $v_3,v_2,v_1,u_5,u_1,u_2,u_3$ do not induce a $P_7$ in $G$, there is an edge between $\{v_1,v_2,v_3\}$ and $\{u_1,u_2,u_3\}$. 

Recall that by assumption, $v_1,v_3,u_1,u_3$ forms an independent set, and analogously, $v_2u_2 \notin E$.  

Since $c,d,h,v_1,u_2,u_3,u_4$ do not induce a $P_7$ in $G$, we have $v_1u_2 \notin E$.       

Since $c,d,h,v_1,v_2,u_1,u_2$ do not induce a $P_7$ in $G$, we have $v_2u_1 \notin E$.       

Since $c,d,h,v_1,v_2,u_3,u_2$ do not induce a $P_7$ in $G$, we have $v_2u_3 \notin E$.       

Since $c,d,h,v_1,v_2,v_3,u_2$ do not induce a $P_7$ in $G$, we have $v_3u_2 \notin E$.        

Now, $v_3,v_2,v_1,u_5,u_1,u_2,u_3$ induce a $P_7$ in $G$, which is a contradiction.

Thus, Case A is excluded.
 
{\bf Case B}: $N_C(u_5)=\emptyset$, i.e., $u_5v_i \notin E$ for $i \in \{1,\ldots,5\}$. 

Since $v_1,h,d,u_5,u_1,u_2,u_3$ do not induce a $P_7$ in $G$, we have $v_1u_2 \in E$. 

Since $c,d,u_5,u_4,u_3,u_2,v_1$ do not induce a $P_7$ in $G$, we have $v_1u_4 \in E$.      

Now, if $u_4v_3 \notin E$ then $c,d,u_5,u_4,v_1,v_2,v_3$ induce a $P_7$ in $G$, which is a contradiction. 

Thus, $u_4v_3 \in E$, i.e., $(h,v_1,u_4,v_3,v_4)$ is a $C_5$ of type 1 with $u_4 \in L(h)$, and we are done with Case B.  

{\bf Case C}: $N_C(u_5)=\{v_2\}$. 

As in Case B, since $v_1,h,d,u_5,u_1,u_2,u_3$ do not induce a $P_7$ in $G$, we have $v_1u_2 \in E$. 

Since $c,d,h,v_1,u_2,u_3,u_4$ do not induce a $P_7$ in $G$, we have $v_1u_4 \in E$.      

Since $c,d,u_5,v_2,v_1,u_2,u_3$ do not induce a $P_7$ in $G$, we have $v_2u_3 \in E$.  

Since $c,d,h,v_4,v_3,v_2,u_3$ do not induce a $P_7$ in $G$, we have $v_4u_3 \in E$.  But then $c,d,u_5,u_4,u_3,v_4,v_3$ induce a $P_7$ in $G$ if $v_3u_4 \notin E$,
 or else $v_3u_4 \in E$ and thus $u_4 \in L(h)$.  

Thus, we are done with Case C.  

{\bf Case D}: Assume by symmetry that $N_C(u_5)=\{v_2,v_4\}$. 

As in Case B and C, since $v_1,h,d,u_5,u_1,u_2,u_3$ do not induce a $P_7$ in $G$, we have $v_1u_2 \in E$. 

Since $c,d,h,v_1,u_2,u_3,u_4$ do not induce a $P_7$ in $G$, we have $v_1u_4 \in E$.      

Since $c,d,u_5,v_2,v_1,u_2,u_3$ do not induce a $P_7$ in $G$, we have $v_2u_3 \in E$.  

Since $c,d,h,v_4,v_3,v_2,u_3$ do not induce a $P_7$ in $G$, we have $v_4u_3 \in E$. But then $c,d,u_5,v_4,u_3,u_2,v_1$ induce a $P_7$ in $G$  which is a contradiction. 
 
Thus, we are done with Case D, and Fact \ref{fact5} is shown.  
\qed

\begin{fact}\label{fact7}
Let $h \in H$ be the nail of a $C_5$ of type $1$ in $K$. Then MWIS can be solved in polynomial time for $G[V(K) \setminus N(h)]$.
\end{fact}

{\bf Proof.} MWIS can be solved for $G[V(K) \setminus N(h)]$ by solving MWIS 
\begin{enumerate}
\item[$(i)$] in $G[V(K) \setminus (N(x) \cup N(h))]$ for any $x \in L(h)$, and 
\item[$(ii)$] in $G[V(K) \setminus (N(h) \cup L(h))]$. 
\end{enumerate}

Note that the subgraphs $G[V(K) \setminus (N(x) \cup N(h))]$ for $x \in L(h)$ contain no $C_5$ of type 1 by Fact \ref{fact5} $(i)$, and that subgraph $G[V(K) \setminus (N(h) \cup L(h))]$ contains no $C_5$ of type 1 by Fact \ref{fact5} $(ii)$ and by definition of $L(h)$. Then steps $(i)-(ii)$ can be executed in polynomial time by referring to Case 1.1.   
\qed 

Then in Case 1.2, MWIS can be solved for $K$ in polynomial time as follows:

Let $A = \{a \in H: a$ is the nail of a $C_5$ of type 1 in $K\}$. Then MWIS can be solved for $K$ by successively solving MWIS 

\begin{enumerate}
\item[$(i)$] in $G[V(K) \setminus N(a)]$ for $a \in A$; 
\item[$(ii)$] in $G[V(K) \setminus A]$ (which contains no $C_5$ of type 1). 
\end{enumerate}

In particular, step $(i)$ can be executed in polynomial time by Fact \ref{fact7}, and step $(ii)$ can be executed in polynomial time by referring to Case 1.1. 

\subsection{Case 2: $G[Z]$ is not bipartite.}

By Lemma \ref{P7K3C5frnearlybip}, we can focus on $C_5$ for odd cycles in $G[Z]$. 

\begin{fact}\label{fact8}
If $G[Z]$ contains a $C_5$ then there is exactly one component of $G[Z]$ which contains a $C_5$.
\end{fact}

{\bf Proof.} First let us observe that since $G$ is ($P_7$,triangle)-free, if a vertex $h \in H$ contacts a component $T$ of $G[Z]$ containing a $C_5$ then $h$ contacts every $C_5$ in $T$; in particular, $h$ is the endpoint of a $P_4$ formed together with three vertices of $T$.

Then assume to the contrary that there are two components, say $Q_1$ and $Q_2$, of $G[Z]$ each one containing a $C_5$. Let $h_1 \in H$ contact $Q_1$. Then by the above, $h_1$ does not contact $Q_2$, since otherwise a $P_7$ arises. Then let $h_2 \in H$ contact $Q_2$ (with $h_2 \neq h_1$). Similarly, $h_2$ does not contact $Q_1$. Then three vertices of $Q_1$, vertices $h_1,d,h_2$, and one vertex of $Q_2$ induce a $P_7$ which is a contradiction. 
\qed 

According to Fact \ref{fact8}, let $Z^*$ be the unique component of $G[Z]$ which is not bipartite, and let $H^* = \{h \in H:h$ contacts $Z^*\}$.

\begin{fact}\label{fact9}
For every $h \in H^*$, $G[Z^* \setminus N(h)]$ is bipartite.
\end{fact}

{\bf Proof.} In fact, since $G$ is ($P_7$,triangle)-free, $h$ contacts every $C_5$ or $C_7$ in $G[Z^*]$.  
\qed 

Then in Case 2, MWIS can be solved for $K$ in polynomial time as follows:

According to the notation above, MWIS can be solved for $K$ by successively solving MWIS
\begin{enumerate}
\item[$(i)$] in $G[V(K) \setminus N(h)]$ for every $h \in H^*$; 
\item[$(ii)$] in $G[V(K) \setminus H^*]$.
\end{enumerate}

Concerning step $(i)$: It can be executed in polynomial time by Facts \ref{fact8} and \ref{fact9}, i.e., by referring to Case 1.

Concerning step $(ii)$: According to Fact \ref{fact8}, $G[V(K) \setminus H^*]$ is partitioned into components, namely $Z^*$ and (possibly) other components $Q$ such that $G[Q \cap Z]$ is bipartite.   

Concerning $G[Z^*]$, MWIS can be solved in polynomial time for $G[Z^*]$ as follows:
\begin{enumerate}
\item[$(a)$] fix any vertex $h^* \in H^*$
\item[$(b)$] solve MWIS for $G[Z^*]$ by referring to Case 1.
\end{enumerate}
In fact, $Z^*$ can be partitioned into independent set $Z^* \cap N(h^*)$ and $Z^* \setminus N(h^*)$ (by Fact \ref{fact9}, $G[Z^* \setminus N(h^*)]$ is bipartite). Concerning the other components $Q$, MWIS can be solved in polynomial time for $G[Q]$ by referring to Case 1.  
\qed

Summarizing the above results, we have:

\begin{theo}
The MWIS problem can be solved in polynomial time for $(P_7$,triangle$)$-free graphs. 
\end{theo}

\section{Concluding remarks}

Recently, Lokshtanov et al. \cite{LokVatVil2014} proved that MWIS can be solved for $P_5$-free graphs in polynomial time, and Lokshtanov et al. \cite{LokPilvan2015} proved that MWIS can be solved for $P_6$-free graphs in quasi-polynomial time $n^{{\cal O}(\log^2n)}$. A natural open problem is whether MWIS can be solved for $P_k$-free graphs in polynomial time for $k \geq 6$ or in quasi-polynomial time for $k \geq 7$ $-$ even though some characterizations of $P_k$-free graphs and some progress are known in the literature \cite{CamSch2015}.

In this manuscript, we show that MWIS can be solved for ($P_7$,triangle)-free graphs in polynomial time (the time bound of our solution algorithm may be estimated as ${\cal O}(n^{13})$). 

This extends the corresponding result for ($P_6$,triangle)-free graphs and may provide some progress in the study of MWIS for $P_7$-free graphs - recall that MWIS remains NP-hard for triangle-free graphs \cite{Polja1974}. It is an open problem whether there is a $k$ such that MWIS is NP-complete for ($P_k$,triangle)-free graphs.

Our approach is based on a repeated application of the anti-neighborhood approach which is based in detail on the simple observation that for any vertex $v$ of a graph $G$ one has $\alpha_w(G)$ = max $\{\alpha_w(G[V(G) \setminus \{v\}]),\alpha_w(G[V(G) \setminus N(v)])\}$. That allows, by detecting an opportune sequence of vertices, to split and to finally reduce the problem to instances of bipartite subgraphs, for which the problem can be solved in polynomial time \cite{AhuMagOrl1993,DesHak1970,GroLovSch1988}. In particular, by the solution method introduced in Section 3, it is not difficult to derive the following result:

\begin{theo}
For every $(P_7$,triangle$)$-free graph $G$ there is a family ${\cal S}$ of subsets of $V(G)$ inducing bipartite subgraphs of $G$, with ${\cal S}$ detectable in polynomial time and containing polynomially many members, such that every maximal independent set of $G$ is contained in some member of ${\cal S}$.   
\end{theo}

The main result of this paper can be extended in various ways as follows:

{\em Remark $1$}. Let us recall two results of Olariu: 
\begin{enumerate}
\item[$(i)$] Every paw-free graph is either triangle-free or complete multipartite \cite{Olari1988}. 
\item[$(ii)$] If a prime graph contains a triangle then it contains a house, bull, or double-gem \cite{Olari1990}. 
\end{enumerate}

Recall that a graph is {\em prime} if it admits no proper (non-trivial) vertex subset $U$ such that all vertices of $U$ are adjacent to the same vertices outside of $U$. Then the result of this manuscript implies that MWIS can be solved for ($P_7$, paw)-free graphs in polynomial time directly by $(i)$, and that more generally MWIS can be solved for ($P_7$, house, bull, double-gem)-free graphs in polynomial time by $(ii)$ and by results from modular decomposition theory (see e.g. \cite{BraLeSpi1999,McCSpi1999,MoeRad1984/1}).  

{\em Remark $2$}. The {\em Minimum Weight Independent Dominating Set} ({\em MWIDS}) problem is the following: Given a graph $G$ and a weight function $w$
on $V(G)$, determine a maximal independent set of $G$ of minimum weight.
The {\em Maximum Weight Induced Matching} ({\em MWIM}) problem is the following: Given a graph $G$ and a weight function $w$ on $E(G)$, determine an induced matching of $G$ of maximum weight (where an {\em induced matching} of $G$ is a matching $M$ of $G$ such that the vertices in $M$ induce a subgraph of $G$ with maximum degree 1).
Since the solution method introduced here to solve MWIS for ($P_7$,triangle)-free reduces iteratively the problem to instances of $P_7$-free bipartite graphs, if MWIDS (if MWIM) can be solved for $P_7$-free bipartite graphs in polynomial time, then MWIDS (then MWIM) can be solved for ($P_7$,triangle)-free graphs in polynomial time. Let us mention that $P_7$-free bipartite graphs have unbounded clique-width \cite{LozVol2008}, and that both MWIDS and MWIM remain NP-hard for bipartite graphs, see respectively \cite{CorPer1984,StoVaz1982}.

{\bf Acknowledgment.} We are grateful for the comments of Fr\'ed\'eric Maffray and Lucas Pastor.

\begin{footnotesize}
\renewcommand{\baselinestretch}{0.4}

\end{footnotesize}

\end{document}